\documentstyle[aps,prl,multicol,epsfig]{revtex}
\begin{document}
\draft
\title{Internal Anisotropy of Collision Cascades}
\author{F. Kun\footnotemark[1]\footnotetext{Electronic address:
    feri@dtp.atomki.hu}, G. Bardos}
\address{Department of Theoretical Physics, Kossuth Lajos University 
         P.O. Box 5, H-4010 Debrecen,  Hungary}

\date{\today}

\maketitle

\begin{abstract}
It is shown that for collision cascades the global fractal
  dimension 
cannot give an adequate description of the geometrical structure
because it is insensitive to the internal
anisotropy of the object arising from the directionality of cascade
branches. In order to give a more elaborate description of the cascade
we introduce an angular correlation function, which takes into account
the direction of the local growth of the branches of the cascades. 
It is demonstrated that the angular correlation function 
gives a quantitative description of the directionality and the
interrelation of branches of the cascade.
The power law decay of the angular correlation is evidenced and
characterized by an exponent $\beta$ and an angular correlation
length $R_a$ different from the radius of gyration $R$. 
It is demonstrated that the overlapping of subcascades
has a strong effect on the angular correlation.

\end{abstract}


\begin{multicols}{2} 
\narrowtext


\section{Motivations}

Collision cascades develop in condensed matter as a consequence 
of irradiation with energetic beams of particles.  The bombarding
particles  transfer 
their kinetic energy in series of collisions with the target atoms
and the energized, recoiling  atoms generate further recoils in their
own slowing-down process.
The result of this energy sharing  process is a collision
cascade. The cascade can also be considered as a far-from equilibrium
process leading to structure formation in the solid target.
Recently the geometrical structure of collision cascades has
been analyzed by means of analytical approaches and Monte Carlo
simulations concentrating on the possible fractal and multifractal
aspects. These investigations
have been extended to the study of the self-similarity properties of
the cascade \cite{cheng1,cheng2,wag},
to the determination of its fractal dimension and multifractal spectra
for different  
interaction potentials \cite{wint,moreno,kun,kun2,web,rossi}, 
and they gave an insight into the cascade-subcascade transition and
into the spike creation \cite{cheng1,cheng2,kun2,rossi}. 

Far from-equilibrium processes often create complex geometrical structures
which exhibit fractal properties characterized by the fractal 
dimension $D$ \cite{bunde}.  
However, the fractal dimension is
a global property of a cluster of particles, it does 
not provide an insight into the structural details of the object.
It has been demonstrated in the case of Diffusion Limited Aggregation
(DLA) that large branching structures  
may be internally anisotropic. This anisotropy results in a tangential
correlation different from the radial one \cite{meakin} and it shows
up in the 
behavior of the three point correlation function as well \cite{halsey}.

The internal structure of collision cascades, the
correlation of the vacancies, the interrelation of branches and the
directionality of ion tracks can have
an important impact on several physical processes, e.g. ion-beam
mixing \cite{cheng3}, radiation enhanced diffusion \cite{cheng3,will}, 
electric and optical properties of irradiated  
polymeric materials \cite{calcagno1,calcagno2}, the development of
mechanical stress due to ion bombardment and crack propagation in
the solid from the irradiated zone \cite{stress}. 

In the case of diffusion, for example, the correlation of 
vacancies due to the directionality of the cascade branches can
enhance the diffusion process in the irradiated zone. 
Inside the cascade the path of the displaced atoms 
can be viewed as conduit, forming a labyrinth in the damaged material
\cite{will}.
Thus, we are led to de Gennes's model of ``termite diffusion'', when
the diffusing particles (termites) meet the conduit-labyrinth, they
diffuse much more quickly and over much greater distances than they do
in the bulk \cite{degennes1}.

Further motivations to study the directionality of ion tracks in a
solid comes from the earlier studies of system of particles
interacting via elastic collisions. A system where many particles
scatter, interacting elastically with each other, may be regarded as a
model of gas.  In such a system each particle is scattered
successively, so it walks almost randomly. Although the velocity
autocorrelation of such a particle was believed to decay
exponentially, Alder and Wainwright discovered a long time tail by
computer simulation \cite{alder1,alder2}:
\begin{eqnarray}
  \label{autocor}
  \left< \vec{v}(t) \cdot \vec{v}(0) \right> \sim t^{-d/2}, \hspace{1cm} d
  \geq 2,
\end{eqnarray}
where $d$ is the dimension of the embedding space.
Long time tails of random walks are also found in a much simpler
system called the Lorentz gas model where a classical particle is
scattered elastically off randomly located fix scatterers. (This model
was originally introduced as a model of electron motion in a metal.)
In the case where the scatterers form a regular lattice, it is proved
that the velocity autocorrelation  has exponential decay. On the
contrary if the scatterers are distributed at random forming a fractal
with dimension $D$ there is a long time tail for the negative values
of the scalar product of the velocities \cite{lorentz}:
\begin{eqnarray}
  \label{autocor1}
  \left< \vec{v}(t) \cdot \vec{v}(0) \right> \sim -t^{-(\frac{d-D}{2}+1)}. 
\end{eqnarray} 
Hence the memory effect in the directionality seems to be a common
feature of the motion of particles scattered randomly. 

Based on the above motivations, in the present paper we want to
obtain a more elaborate picture of  
collision cascades
gaining information about the internal 
branching morphology. 
The computer simulations have the advantage that one can monitor 
such quantities
which are hard to measure or are not measurable but they can help the
understanding of a physical process. 
In the study of ion-solid interaction in certain cases not only the
distribution of the vacancies is of interest but even the path of the
moving particles. Obviously, the study of the ions's paths includes  
the vacancies as well, the position of the vacancies are the branching
points in the cascade tree.
Thus in the present study not only the
vacancies but 
all the collisional points are included in the analysis of the cascade
regardless of the vacancy creation, that can easily be done by means
of computer simulations. We performed Monte Carlo simulations of self
ion collision cascades in two
and three dimensions. For the three-dimensional simulations the
scattering cross section $d\sigma = C(m)E^{-m}T^{-1-m}dT$ belonging to
the inverse power law potential $V(r) \sim r^{-1/m}$ was applied. Here
$E$ is the energy of the bombarding particle, $T$ is the energy
transferred to the target atom in the scattering process and 
$m$ denotes the parameter of the interaction potential $0 < m \leq 1$.
In two dimensions we established a toy model with the scattering cross
section $d \sigma = KE^{-\frac{m}{2}}T^{-1-\frac{m}{2}}dT$
\cite{kun2}. A two 
dimensional cascade does not have experimental relevance but it can
help to reveal the role of the embedding dimension $d$ in the
formation of the cascade structure. For further details about computer
simulation of 
ion-solid interaction see also Ref. \cite{eckst}.

In the following, at first the structure of cascades generated by
Monte Carlo simulations is
analyzed in terms of the density-density correlation function. Then
an angular correlation function is introduced based on
the direction of the local growth of the cascade branches.
It is shown that the angular correlation function 
gives a quantitative description of the directionality and the
interrelation of branches of cascades, furthermore, the role of the
intersection of branches is enlightened.

\section{Correlation Functions}
\begin{figure}
\begin{center}
\epsfig{bbllx=19,bblly=390,bburx=510,bbury=605,file=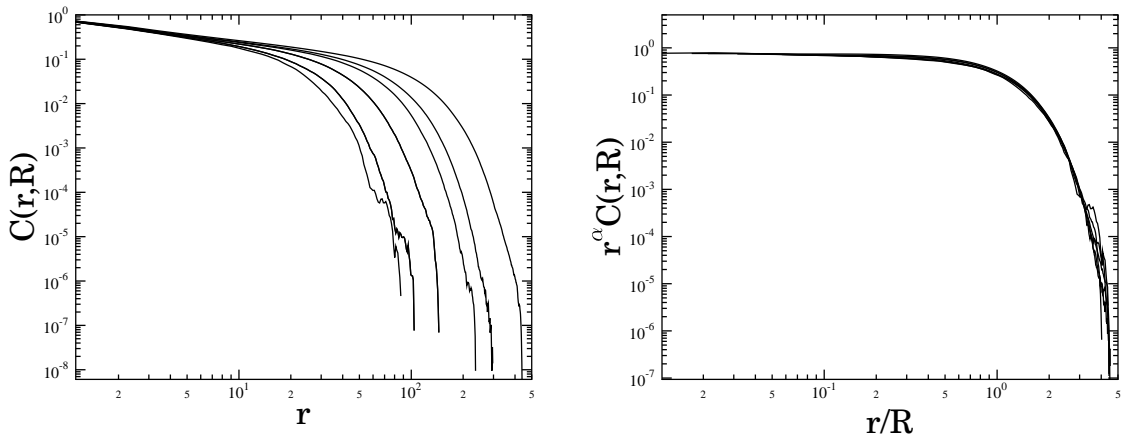,
width=15cm}
\caption{The density-density correlation function for system sizes
  from $N=400$ to $N=4000$ in the $2d$ model. The parameter $m$ of the
  scattering cross section  was chosen to be $m=0.7$. The data
  collapse analysis was performed with ten different curves.}
\label{fig:corr2d}
\end{center} 
\end{figure}

\begin{figure}
\begin{center}
\epsfig{bbllx=19,bblly=390,bburx=510,bbury=605,file=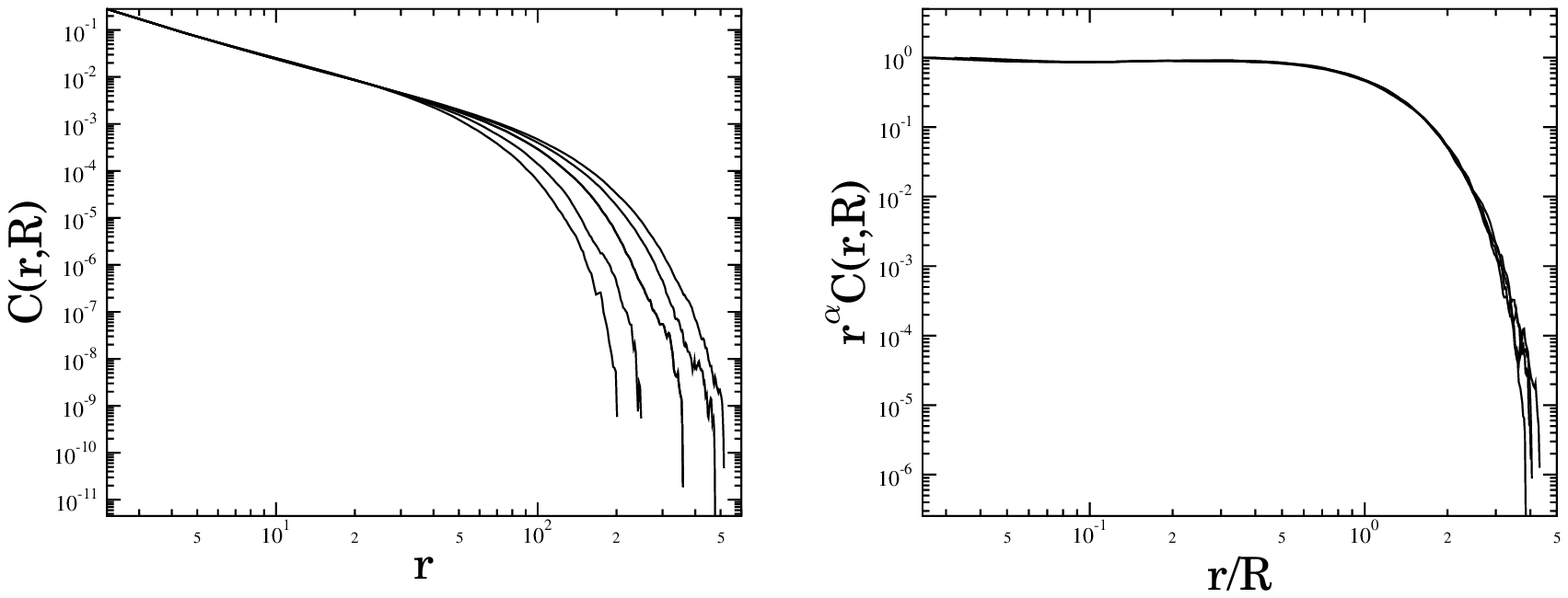,
width=15cm}
\caption{ The density-density correlation function for system sizes
  from $N=800$ to $N=5000$ in the $3d$ model. The parameter $m$ of the
  scattering cross section  was chosen to be $m=0.4$. The data
  collapse analysis was performed with ten different curves.}
\label{fig:corr3d}
\end{center} 
\end{figure}

The usual way of studying the structure of an object composed of $N$ particles
in the $d$ dimensional embedding space
is in terms of the density-density correlation function $C(r)$:
\begin{eqnarray}
C(r)=\frac{1}{Nr^{d-1}\Omega_d \delta r}
 \sum_{r-\frac{\delta r}{2} < |\vec{r}_i-\vec{r}_j|<r+\frac{\delta r}{2}} 
\rho( \vec{r}_i) \rho( \vec{r}_j),
\label{eq:corr}  
\end{eqnarray}
where $\rho(\vec{r})$ is $1$ if there is particle at $\vec{r}$ and zero
otherwise. In our
case these particles are the points where collisions occurred in the
target material during the evolution of the cascade.
$\Omega_d$ denotes the solid angle in $d$ dimension. 
For structures of finite size, like collision cascades, the two point
density-density
correlation function depends on the overall size of the structure
described by a characteristic length $R$ as well as the internal
length $r$. In this case the correlation function can be written as
$C(r,R)$. Here $C(r,R)$ represents the correlation function averaged
over a large number of cascades of the same size $R$.
The characteristic macroscopic length of the cascade is 
the radius of gyration $R$ defined as the average distance 
of the particles:
\begin{eqnarray}
R^2 = \frac{1}{N^2}\sum_{i,j} |\vec{r}_i-\vec{r}_j|^2 \sim
\frac{\Omega_d}{N}\int drr^{d+1}C(r) 
\end{eqnarray}
$C(r,R)$ is normalized by the total number of particles $N$:
\begin{eqnarray}
N=\Omega_d \int drr^{d-1}C(r) 
\label{eq:n}
\end{eqnarray}
Assuming that $C(r,R)$ is a homogeneous function of its variables
results in the scaling form:
\begin{eqnarray}
C(r,R) = r^{-\alpha}g(x), \hspace{1.5cm}   x=\frac{r}{R}, 
\label{eq:scal}
\end{eqnarray}
where $g(x)$ is constant for $x<<1$ and $g(x) \sim e^{-x}$ for $x > 1$. 
The function $g(x)$ is called scaling function and the exponent
$\alpha$ is the scaling exponent.
The fractal dimension $D$ of the object is defined through the
behavior of the total number
of particles $N$ as a function of the radius of gyration $R$, i.e. $N \sim
R^D$. From Eqs. (\ref{eq:n},\ref{eq:scal}) it follows that $\alpha =
d-D$. If Eq. (\ref{eq:scal}) provides a valid description of the
geometric scaling properties then plots of $r^{\alpha}C(r,R)$
vs. $r/R$ for structures of different sizes will fall on a common
curve (on the scaling function $g(x)$). 
This data collapse provides a
reliable way of measuring the scaling exponent $\alpha$ (and thus the fractal
dimension $D$) by varying the value 
of $\alpha$ until the best collapse is obtained.
Representative examples of the density-density correlation function
and its data collapse analysis can be seen in
Figs. \ref{fig:corr2d}, \ref{fig:corr3d} for 
the $2d$ and $3d$ models. The excellent collapse obtained with ten different
system sizes proves the validity of the scaling ansatz of
Eq. (\ref{eq:scal}) for both model systems. 

\begin{figure} 
\begin{center}
\epsfig{bbllx=46,bblly=116,bburx=445,bbury=516,file=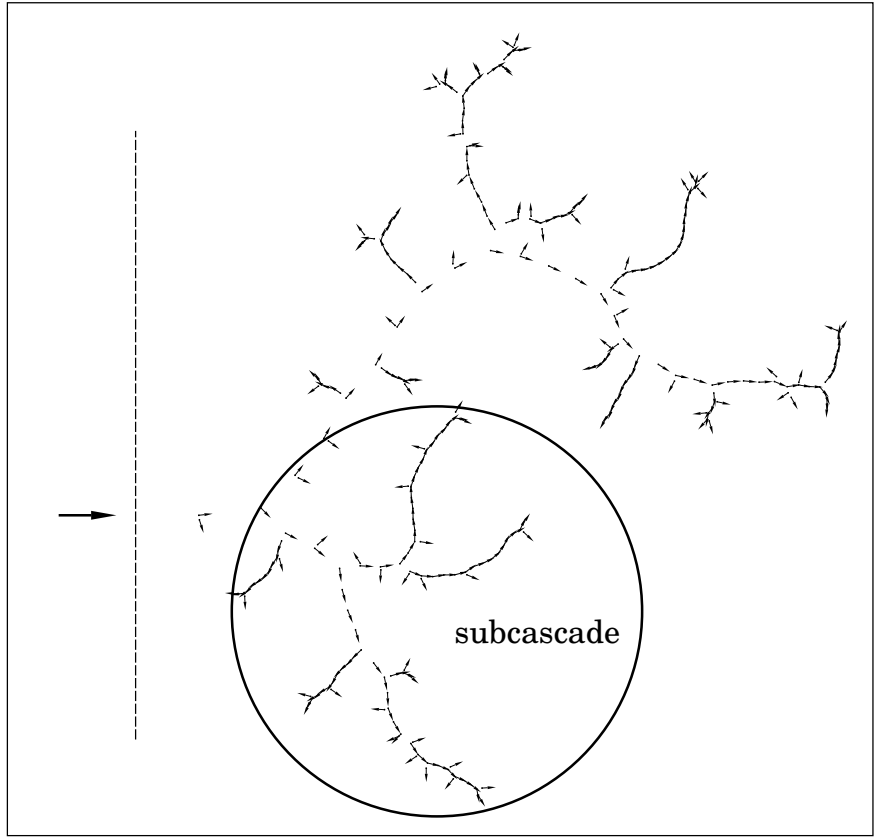,
width=12cm}
\caption{ Collision cascade generated in $2d$. The arrows represent the
unit vector $\vec{p}(\vec{r}_i)$ of the linear momentum 
of the scattered particles after the 
collision. $m$ has the same value as in Fig. \ref{fig:corr2d}. 
The circle indicates an extended subcascade resulted from a high
energy recoil. The surface of the solid and the direction of the
penetration of the bombarding particle are also shown.} 
\label{fig:cas}
\end{center} 
\end{figure}

However, this way of description based on $C(r,R)$ cannot
reveal anything about the possible internal anisotropy because the
density-density correlation function is insensitive to the structural
details. 
In order to obtain information about the branching structure
we have to take into account some dynamical features of the growth of
the cascade. For this purpose
we assign to each particle, i.e. to each point of 
collision, with position $\vec{r}_i$ the unit vector $\vec{p}(\vec{r}_i)$ of
the linear momentum of the scattered particle.
If there is vacancy creation in a collision
the unit vector of the recoiled particle appears at the position of
its first collision. With this prescription one and only one unit
vector is assigned to each point of the object studied.
The
$\vec{p}(\vec{r}_i)$ vector characterizes the direction of the local growth
of the corresponding branch of the cascade tree at position $\vec{r}_i$ 
during the cascade evolution. Fig. \ref{fig:cas} shows the unit
vectors $\vec{p}$ attached to the particles.
Using the 
vector field $\vec{p}(\vec{r}_i)$  an angular correlation function 
$C_a(r,R)$ can be introduced with the following definition:
\begin{eqnarray}
C_a(r) = 
\frac{1}{N r^{d-1} \Omega_d \delta r}
 \sum_{r-\frac{\delta r}{2} < |\vec{r}_i-\vec{r}_j|<r+\frac{\delta r}{2}} 
\vec{p}( \vec{r}_i) \cdot \vec{p}( \vec{r}_j)
\label{eq:acorr}
\end{eqnarray}
The form of this equation is similar to Eq. (\ref{eq:corr}) but in the
summation the scalar product of the $\vec{p}(\vec{r}_i)$ 
vectors is used
instead of the product of the one particle densities $\rho(\vec{r}_i)$.

\begin{figure}[!h]
\begin{center}
\epsfig{bbllx=47,bblly=390,bburx=371,bbury=541,file=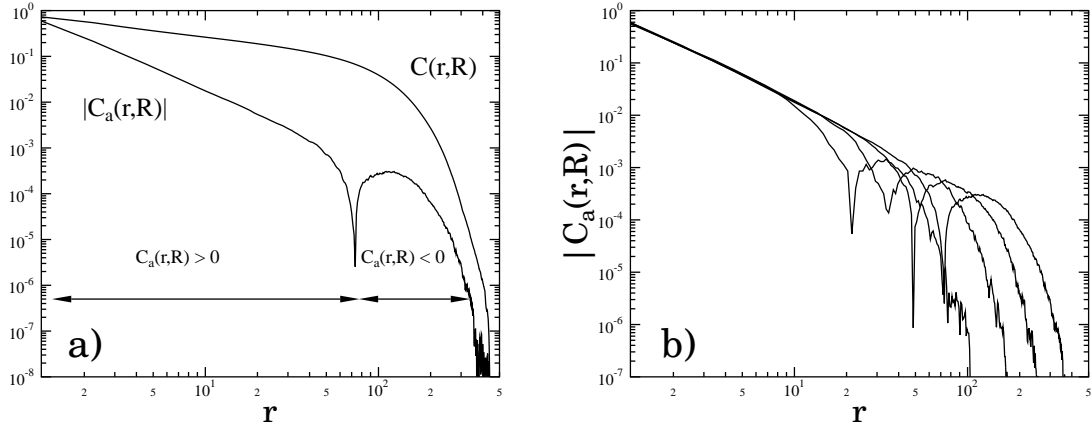,
width=15cm}
\caption{ Representative examples of the behaviour of the
  angular correlation function $C_a(r,R)$ in the same system as in
  Fig. \ref{fig:corr2d}. $(a)$ Comparison of the
  absolute value of the angular correlation function $|C_a(r,R)|$ and
  the density-density correlation function $C(r,R)$ belonging to the
  same system size $R$, $(b)$ $|C_a(r,R)|$ for different system sizes. }
\label{fig:acorr}
\end{center} 
\end{figure}

Along a given branch of the cascade the vectors $\vec p(\vec{r}_i)$
are correlated in 
the sense that they are almost 
parallel to each
other hence the scalar product in (\ref{eq:acorr}) has values close to
$1$. Subbranches appear in the cascade tree as a result of the
recoiled particles.
In Fig. \ref{fig:cas} it can be observed that a high energy recoil can
give rise even to an extended subcascade the branches of which grow
independently of the other parts of the cascade, the other
subcascades. This implies that the average value of the
scalar product of vectors $\vec{p}(\vec{r}_i)$ belonging to different 
subcascades 
is close to zero or it can even be negative. In this sense the angular
correlation function $C_a(r,R)$ can measure the directionality of
cascade branches with respect to each other.
Representative examples of the absolute value of the angular 
correlation function $|C_a(r,R)|$ of cascades generated in the $2d$
model are shown in Fig. \ref{fig:acorr}.
Fig. $\ref{fig:acorr}a$ presents a comparison of the angular and the
density-density correlation
functions belonging to the same system and
Fig. $\ref{fig:acorr}b$ shows $|C_a(r,R)|$ for different system sizes $R$.
It can be observed that for a certain length range $C_a(r,R)$ decays
according to a power law with an exponent significantly different from
the exponent 
describing the algebraic decay of $C(r,R)$.
At a certain value of $r$ the angular correlation function
becomes negative and it goes to zero through negative values at long
distances.  Let's denote $R_a$ the characteristic length where
 $C_a(r,R)$ becomes negative. 
Increasing the system size $R$ in Fig $\ref{fig:acorr}b$
$R_a$ is also increasing. The negative part of $C_a(r,R)$ has a local
minimum (i.e. $|C_a(r,R)|$ has a local maximum) and it goes to zero
exponentially due to the finite size cutoff.

Based on the above observations,  the same
form of scaling ansatz can be assumed for $C_a(r,R)$  as for  
$C(r,R)$, i.e. Eq. (\ref{eq:scal}):
\begin{eqnarray}
C_a(r,R) = r^{-\alpha_a}f(x), \hspace{1.5cm}   x=\frac{r}{R}, 
\label{eq:scala}
\end{eqnarray}
where $f(x)$ denotes the scaling function and $\alpha_a$ the scaling
exponent belonging to $C_a(r,R)$. 
At small distances $C_a(r,R)$ approaches $C(r,R)$ thus
part of the power law decay of $C_a(r,R)$ can be attributed to the
power law decay of $C(r,R)$, therefore the exponent $\beta$  characterizing 
the angular correlation is the difference of $\alpha_a$ and $\alpha$:
\begin{eqnarray}
  \label{eq:beta}
   \frac{C_a(r,R)}{C(r,R)} \sim r^{-\beta}, \hspace{1cm} 
 \beta = \alpha_a-\alpha
\end{eqnarray}
Finally, the internal anisotropy of cascades  is determined by two
quantities, by the characteristic length $R_a$ where $C_a(r,R)$ drops
to negative values  and the exponent
$\beta$ describing the speed of the decay of  $C_a(r,R)$  . 

In the next section we present the results of the analysis of
systematic Monte Carlo simulations in terms of the correlation
functions introduced above, and we try to give a simple interpretation
of the results obtained.

\section{Results and Discussion}

In a recent publication (see Ref. \cite{kun2}) it has been
demonstrated that the parameter $m$ 
of the interaction potential plays the role of a control parameter
from the viewpoint of the cascade geometry. For decreasing $m$ a
structural transition takes place in the cascade from an open
branching structure to a space filling one. When the upper critical
dimension $d_u$ of the cascade exceeds the dimension of the embedding
space $d$, so called geometrical correlations arise in the cascade due
to the intersection of different branches and the overlapping of
subcascades. These geometrical correlations lead to a fractal
dimension $D(m)$ different from the self similarity dimension $D_o(m)$
for $m \leq 1$
and $m \leq \frac{1}{3}$ in the $2d$ and $3d$ models, respectively
\cite{kun2}.  
The question naturally arises how the intersection of
branches influences the behavior of the angular correlation function
introduced. 

 The above analysis in terms of the density-density and angular correlation
functions was performed for cascades in a wide range of the
parameter  $m$ of the potential ( in the same range as in
Ref. \cite{kun2}) for both model systems. 
The scaling exponents $\alpha, \alpha_a$ and the characteristic 
length $R_a$ were obtained by means
of the data collapse method. 
Since the results concerning to the fractal
dimension of cascades have already been discussed in Ref. \cite{kun2}
here we restrict ourself to the discussion of the
properties of the angular correlation function. 

\begin{figure}[!h]
\begin{center}
\epsfig{bbllx=44,bblly=329,bburx=323,bbury=564,file=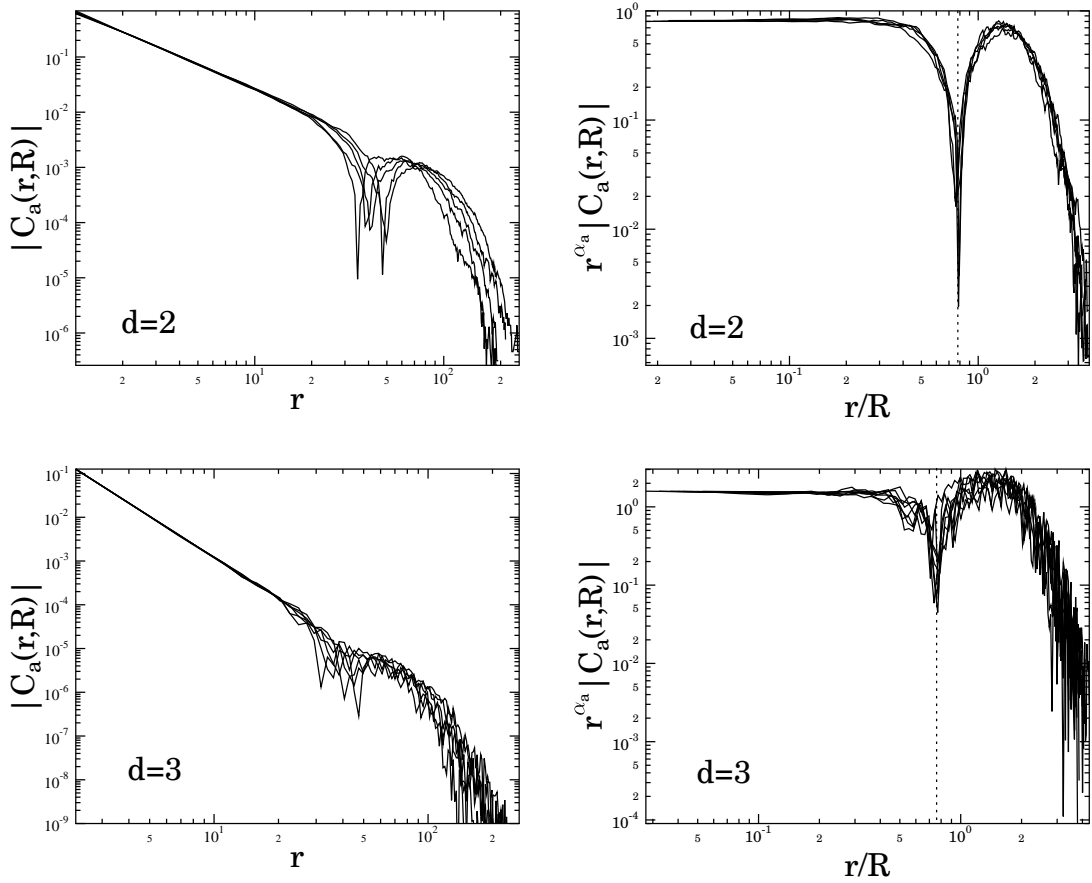,
width=15cm}
\caption{ Data collapse analyzis 
  of the angular correlation function in the $2d$ and $3d$ models. 
  One can observe the good quality of the collapse, which demonstrates the
  validity of the scaling 
  ansatz Eq. (\ref{eq:scala}). The parameter $m$ of the scattering
  cross section was chosen to be $m=0.7$ and $m=0.4$ in $2d$ and $3d$,
  respectively. The value of $\gamma=R_a/R$ is indicated by dotted
  lines in the figures, it is $0.78$ and $0.76$  in $2d$ and $3d$.}
\label{fig:collap2d3d}
\end{center} 
\end{figure}

Fig. \ref{fig:collap2d3d} shows
representative examples of the data collapse analysis of the angular
correlation function in the $2d$ and
$3d$ models. It can be observed that $C_a(r,R)$ exhibits power law
behavior for more than one order of magnitude in length.
The good quality of the collapse obtained
demonstrates the validity of the scaling ansatz Eq. (\ref{eq:scala}).
Note, that the positive and negative parts of $C_a(r,R)$
have the same scaling properties. 
The power law behavior of $C_a(r,R)$ expresses the fact that there is long
range angular correlation inside the cascade, which manifests in
the directionality of the branches. The negative 
regime of $C_a(r,R)$ can be interpreted as 
anti-correlation of the local growth directions at large
distances. This implies that the angle of the unit vectors $\vec p(\vec{r}_i)$,
$\vec p(\vec{r}_j)$  tends to be larger than $\frac{\pi}{2}$ for 
 $|\vec{r}_i-\vec{r}_j| > R_a$. The characteristic
length $R_a$ where $C_a(r,R)$ drops to negative values can be
considered as an 
angular correlation length. Let us denote $\gamma$ the ratio of the angular
correlation length $R_a$ to the radius of gyration $R$.   

From the scaling analysis, the exponent $\beta$ characterizing the
decay of $C_a(r,R)$ and the ratio $\gamma$ of the angular correlation length
$R_a$ to the radius of gyration $R$ can be determined precisely.
From the simulations it was found that $\gamma$ is between $0.7$ and
$0.8$ for both model systems and within the accuracy of the
calculations we could not reveal any systematic
dependence of $\gamma$ on the parameter of the potential $m$.

\begin{figure}[!h]
\begin{center}
\epsfig{bbllx=19,bblly=390,bburx=510,bbury=605,file=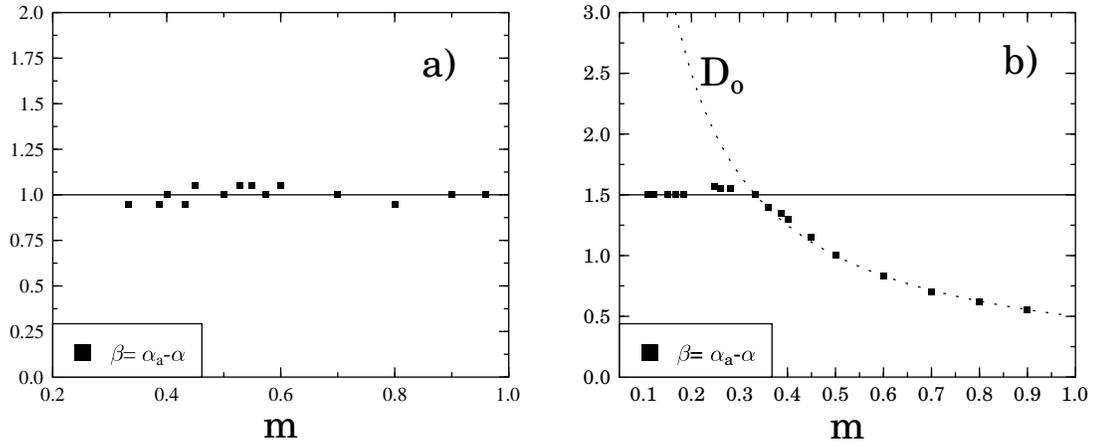,
width=15cm}
\caption{ The exponent $\beta=\alpha_a-\alpha$ characterizing
  the decay of the 
  angular correlation function $C_a(r,R)$ $(a)$ for $2d$ and $(b)$ for
  $3d$. In the $3d$ case the curve of the self similarity dimension is
  also shown. The horizontal solid lines indicate the value of $d/2$.} 
\label{fig:m_iner}
\end{center} 
\end{figure}

Fig. \ref{fig:m_iner} presents the values of $\beta$ as a function of 
$m$ for $2d$ and $3d$. It is
important to note that small value of $\beta$ corresponds to slow decay of the
directionality, i.e. to long range angular correlation, while increasing
$\beta$ implies the weakening of the angular correlation. 
In Fig. $\ref{fig:m_iner}b$ one can observe that in $3d$  two
different regimes of $\beta$ can be distinguished.
If $m \geq 1/3$, when no overlap occurs in the cascade, $\beta$ practically 
coincides with the self
similarity dimension $D_o$. In this region $\beta$ is increasing for
decreasing $m$. When $m$ is close to one (Coulomb-scattering) the
probability of creating a 
high energy recoil is rather small, therefore dense subcascades cannot
occur, the cascade is strongly
directed giving rise to a small value of $\beta$. Decreasing $m$ gives
rise to an increasing number of subcascades growing independently,
which results in weakening
of the directionality order. This is expressed by the increasing value
of $\beta$. At $m=1/3$ the self similarity dimension $D_o$ and $\beta$
reaches $d/2$, half of the dimension of the embedding space.
For $m < 1/3$ when the overlap of the branches dominates the
cascade structure, 
$\beta$ remains practically independent of $m$, 
$\beta = d/2$. 
But this does not entails that the density-density correlation function
$C(r,R)$ and the angular correlation function $C_a(r,R)$ are
independent of $m$. Their exponents $\alpha_a$ and $\alpha$ are
decreasing with decreasing $m$ but in such a way that their difference
remains constant:
\begin{eqnarray}
  \label{beta1}
  \beta= \alpha_a - \alpha = \frac{d}{2}.
\end{eqnarray}
This argument is also supported by the two dimensional simulations
(see Fig. $\ref{fig:m_iner}a$). It has been shown in Ref. \cite{kun2}
that in the two
dimensional model the overlap effect, the intersection of branches is
always present, in the whole region of $m$.
In Fig. $\ref{fig:m_iner}a$ it
can be seen that $\beta$ is equal to $1$, i.e. $\beta =d/2$, as it can
be expected from the $3d$ case, see Fig. $\ref{fig:m_iner}b$. 
The result that $\beta = \frac{d}{2}$ throughout the overlapping
regime demonstrates that the $\vec{p}(\vec{r}_i)$ vector field does
not become isotropic even for high degree of overlap, i.e. for small
values of $m$. In the system
there is always a preferred direction, namely, the initial direction of the
bombarding particle. This also manifests  in the overall shape of the
cascade. The overall shape is not spherical but it is always elongated
along the initial direction of the bombarding particle. 

It has been mentioned in the introductory part 
that the internal anisotropy of DLA clusters was studied by
means of the tangential and the three-point correlation functions,
which are essentially special type of density-density correlation
functions. It is important to mention that, beside these techniques,
Meakin introduced an angular 
correlation function of the type of Eq. (\ref{eq:acorr})
\cite{meakin_ang}. He assigned a so-called bond vector to each
particle in the aggregate. The bond vector $\vec{b}$ is a unit vector
pointing in the direction of the bond, which is formed when a new
particle is added to the growing cluster. Using the $\vec{b}$ vectors an
angular correlation function could be defined. 
$\beta = 0.31$
was found in two dimensional off-lattice simulations \cite{meakin_ang}
that evidenced the slow decay of the
directionality of the cluster branches in DLA processes. 

Collision
cascades have the advantage that the structure formation can be easily
controlled by varying the parameter of the interaction potential $m$.
This gives us a rich spectrum of possibilities to study the
internal anisotropy resulted from the directionality of the branches
of the object and it makes also possible to gain information about the
role of the intersection of branches. 

\section{Conclusions}

In the present paper we studied the internal
anisotropy of collision cascades arising from the branching
structure. It was demonstrated that the global fractal dimension
cannot account for the internal details of the structure. 
To give a more elaborate description of the cascade it was necessary
to introduce an angular correlation function, which takes into account
the direction of the local growth of the cascade branches. 
With the help of the angular correlation function we could quantify
the anisotropy, which manifest in the directionality of the branches. 
The power law decay of the angular correlation was evidenced and
characterized by an exponent $\beta$ and an angular correlation
length $R_a$ different from the radius of gyration $R$. 
The overlapping of subcascades has a strong effect on
the angular correlation. In the absence of overlap $\beta$ coincides
with the self similarity dimension, while in the presence of overlap
$\beta$ is constant, i.e. $\beta = d/2$. It is interesting to note
that in the presence of overlap solely the dimension of the embedding
space determines the speed of the decay of the directionality order.
We argued that this internal anisotropy of growth directions shows up
in the 
overall shape of the cascade as well. 

\section{Acknowledgment}
The authors are very grateful to K. Sailer and Zs. Gul\'acsi for the
valuable discussions. This work was supported by OTKA T-023844.

\end{multicols}
\widetext

\section{Figures}

\end{document}